# Possible quadrupole order in tetragonal Ba$_2$CdReO$_6$ and chemical trend in the ground states of 5$d^1$ double perovskites

Daigorou Hirai[*] and Zenji Hiroi

Institute for Solid State Physics, University of Tokyo, Kashiwa, Chiba 277-8581, Japan

*E-mail: dhirai@issp.u-tokyo.ac.jp

**Abstract**

The synthesis and physical properties of the double perovskite (DP) compound Ba$_2$CdReO$_6$ with the 5$d^1$ electronic configuration are reported. Three successive phases originating from a spin–orbit-entangled $J_{\text{eff}}$ = 3/2 state, confirmed by the reduced effective magnetic moment of 0.72 $\mu_B$, were observed upon cooling. X-ray diffraction measurements revealed a structural transition from a high-temperature cubic structure to a low-temperature tetragonal structure at $T_s$ = 170 K, below which the $J_{\text{eff}}$ = 3/2 state was preserved. Magnetization, heat capacity, and thermal expansion measurements showed two more electronic transitions to a possible quadrupole ordered state at $T_q$ = 25 K, and an antiferromagnetic order of dipoles accompanied by a ferromagnetic moment of ∼ 0.2 $\mu_B$ at $T_m$ = 12 K. These properties were compared with those of the sister compounds Ba$_2$BReO$_6$ (B = Mg, Zn, and Ca) and the chemical trend is discussed in terms of the mean-field theory for spin–orbit-coupled 5$d$ electrons [G. Chen *et al.*, Phys. Rev. B **82**, 174440 (2010)]. The DP compound Ba$_2$BReO$_6$ provides a unique opportunity for a systematic investigation on symmetry breaking in the presence of multipolar degrees of freedom.

Keywords: spin–orbit interaction, multipolar order, double perovskite, Ba$_2$CdReO$_6$,

## 1. Introduction

A remarkable variety of electronic orders associated with the spin and orbital degrees of freedom occurs in solids; time-reversal symmetry breaking causes spins alignments in ferromagnetic (FM) or antiferromagnetic (AF) orders, and orbitals are coupled with crystal lattices to cause a regular arrangement of selected orbitals. In most 3$d$ electron systems, these two degrees of freedom are almost independent and give rise to complex orders [1–4], while, in compounds with heavy elements, they are entangled into a multipolar degree of freedom owing to relatively strong spin–orbit interactions (SOIs). Thus, exotic multipolar orders beyond the conventional spin and orbital orders can occur in compounds containing heavy elements.

Multipolar orders have been studied mainly in $f$-electron based lanthanide and actinide compounds, in which localised $f$-electrons have unquenched orbital angular momenta [5–7]. Recently, 4$d$ and 5$d$ electron systems with moderate SOIs have been the focus as novel platforms for multipolar physics. A $t_{2g}$ manifold of a $d$ electron in an octahedral crystal field possesses an effective angular momentum ($L_{\text{eff}}$) of 1, which is entangled with a spin angular momentum by SOIs to give an electronic state characterised by the total angular momentum $J$. For example, a $J_{\text{eff}}$ = 1/2 state was confirmed in iridates [8,9], and other spin–orbit-entangled states were explored for other electron counts [10,11,20,12–19] and molecule-based $J$ states [21–23]. Among them, octahedrally coordinated transition metal (TM) ions with the $d^1$ and $d^2$ electron counts play a central role in searching for multipole orders because of their highly degenerate ground states. The combination of a high-symmetry crystal field and strong SOIs stabilises a $J$ = 3/2 quartet and a $J$ = 2 quintet as ground states for the $d^1$ and $d^2$ electronic configurations, respectively [24]. In the presence of multipolar degrees of freedom, orderings of quadrupole and octupole moments are proposed [10–15]. Compared to the $f$ electron systems, the spatially extended nature of $d$ orbitals enhances direct interactions between multipoles to stabilise unique multipolar orders in 4$d$/5$d$ electron systems.

Double-perovskite (DP) compounds [25] provide a material platform hosting a wide variety of multipolar orders. They have the chemical formula $A_2BB'O_6$, where $B$ and $B'$ denote a nonmagnetic ion and a 4$d$ or 5$d$ TM ion, respectively. The octahedrally coordinated $B$ and $B'$ ions form a rock salt-type arrangement in the cubic DP structure with the space group $Fm\text{–}3m$. DP compounds are mostly Mott insulators because of the relatively large distances between TM ions, although most of the 5$d$ TM compounds are weakly correlated metals with large bandwidths. Furthermore, the regular octahedral coordination at the TM site results in an unquenched orbital moment that is entangled with spin to give multipolar degrees of freedom.

Signatures of quadrupole and octupole orders have been observed in various DP compounds [26–34]. Among them, quadrupole ordering was intensively discussed in $Ba_2NaOsO_6$, which contains $Os^{7+}$ ions with the 5$d^1$ electronic configuration [26–29]. Nuclear magnetic resonance (NMR) studies revealed an unusual noncollinear AF structure accompanied by a large FM moment induced by spin canting with an angle of 67° below $T_N$ = 6.8 K [26]. Theoretical work suggested that this magnetic order is driven by a quadrupole order existing at higher temperatures [10,35]. Indeed, the NMR measurements observed breaking of the fourfold lattice symmetry while preserving the time-reversal symmetry between 10 and 11 K in an applied magnetic field of 9 T [26]. Moreover, a phase transition, possibly originating from the quadrupole order, was observed at 9.5 K at zero field in recent heat capacity and magnetic susceptibility measurements [27]. On the other hand, in another DP $Ba_2MgReO_6$ containing $Re^{6+}$ ions with the 5$d^1$ configuration, a quadrupole order and a noncollinear AF order with a spin canting angle of 40° were observed by bulk material properties and diffraction measurements [30–32]. For the 5$d^2$ DPs $Ba_2BOsO_6$ ($B$ = Zn, Mg, and Ca) containing $Os^{6+}$ ions, the possibility of a ferro-octupole order was proposed to explain the experimental observations [33]: the time-reversal symmetry breaking observed in muon spin relaxation experiments, the absence of a magnetic Bragg peak in neutron diffraction measurements, and the spin gap observed in inelastic neutron scattering experiments.

Although several DPs exhibit similar multipolar orders, different ground states sometimes appear in spite of the same electronic count and crystal structure. For example, $Ba_2LiOsO_6$, which is isostructural and isoelectronic with $Ba_2NaOsO_6$, shows only one phase transition at 8 K to an AF order without an FM moment [28]. The origin of the different ground states has hereto not been understood. To unveil a critical factor that determines multipolar ground states, a systematic study with varying of specific parameters is highly desirable.

The DP family $Ba_2BReO_6$ containing $Re^{6+}$ (5$d^1$) provides an ideal platform for such a systematic study of multipolar ground states, thanks to the wide variety of divalent cations on the $B$ site. There are three cubic DP compounds in addition to $Ba_2MgReO_6$: $Ba_2ZnReO_6$ [36], $Ba_2CdReO_6$ [37,38], and $Ba_2CaReO_6$ [39,40]. Their lattice constants systematically increase, appearing in the order $Ba_2MgReO_6$ < $Ba_2ZnReO_6$ < $Ba_2CdReO_6$ < $Ba_2CaReO_6$ as the ionic radii of the $B$ site cations increase as Mg < Zn < Cd < Ca. Accordingly, the electronic ground states change systematically. $Ba_2MgReO_6$ and $Ba_2ZnReO_6$ with relatively small lattice constants show FM behaviour, as evidenced by hysteresis in the isothermal magnetisation below the magnetic transition temperatures $T_m$ = 18 and 11 K, respectively [30,36]. Above $T_m$, the $B$ = Mg and Zn compounds show additional phase transitions at approximately the same temperature, $T_q$ = 33 K, which was revealed to be a quadrupole ordering transition for $B$ = Mg; for $B$ = Zn it was unclear because the transition was very broad, possibly due to the poor sample quality or cation site mixing. In contrast, $Ba_2CaReO_6$, which has the largest unit cell, shows a single-phase transition to a collinear AF order below $T_m$ = 15.4 K [40]. For $Ba_2CdReO_6$ with an intermediate lattice constant, the magnetic ground state has not been reported.

In this work, we report on the synthesis and physical properties of $Ba_2CdReO_6$. X-ray diffraction (XRD) measurements revealed a transition from a high-temperature cubic structure to a low-temperature tetragonal structure at $T_s$ = 170 K. At $T_q$ = 25 K, another phase transition was observed in heat capacity and linear thermal expansion. Furthermore, an FM behaviour with a saturation moment of 0.21 $\mu_B$ was observed below $T_m$ = 12 K. These successive phase transitions at $T_q$ and $T_m$ are very similar to those observed in $Ba_2MgReO_6$ with the smallest lattice constant. Thus, the $B$ = Mg, Zn, and Cd compounds are likely to share common electronic ground states. Since the magnetic order of the $B$ = Ca analogue is different from the others, there must be a boundary between $B$ = Cd and Ca. It seems that the ground states of $Ba_2BReO_6$ are sensitive to subtle changes in the lattice constant. Understanding the reason behind the same will shed light on the microscopic mechanism of multipolar ordering in 5$d$ electron systems.

## 2. Experimental

Polycrystalline $Ba_2CdReO_6$ samples were prepared by a conventional solid-state reaction. BaO, CdO, and $ReO_3$ powders were mixed at a 2:1:1 ratio in an argon-filled glove box. The mixture was pelletised and sealed in an evacuated quartz tube. The tube was heated at 700 °C for 24 h. The sintered pellet was crushed and re-pelletised in a glove box, and then heated in an evacuated quartz tube at 800 °C for 24 h. The obtained pellet had a dark blue colour.

The synthesised polycrystalline samples were characterised by XRD in a diffractometer (SmartLab, Rigaku Corporation, Japan) with monochromatic Cu K$\alpha$ radiation.

Electrical resistivity measurements were performed with a standard four-probe configuration in a physical properties measurement system (PPMS, Quantum Design, USA). Electrical contacts were made using silver paint (DuPont 4922N). Magnetic susceptibility was measured using a magnetic properties measurement system (MPMS-3, Quantum Design, USA). A polycrystalline pellet was attached to a quartz sample holder with varnish. Heat capacity measurements were made using a semi-adiabatic thermal relaxation technique in the PPMS. Thermal contact between the sample and sapphire sample stage was made using Apiezon-N grease.

The linear thermal expansion, $\Delta L(T)/L(300\ \text{K})$, was measured by the strain gauge method. A strain gauge (KFL05-120-C1-11, gauge factor 2.01 at 297 K, Kyowa, Japan) was glued to a sintered pellet of polycrystalline $Ba_2CdReO_6$, and the thermal variation of the sample size was recorded as a change in the resistance of the strain gauge. The strain gauge was calibrated by measuring a copper plate (purity 99.99%) with a known linear thermal expansion [41]. Because we used polycrystalline samples, the measured $\Delta L/L$ was approximately proportional to the expansion of volume $V$ in the relation $\Delta V/V = 3\Delta L/L$.

## 3. Results

### 3.1 Structural transition

As shown in Fig. 1a, the XRD patterns of $Ba_2CdReO_6$ measured at 4 and 300 K are significantly different, indicating a structural transition during cooling. The XRD pattern at 300 K is indexed based on a face-centred cubic cell with a lattice parameter of $a = 8.3253(1)$ Å, which is close to the $a = 8.322$ Å previously reported for the powder sample [38]. No site mixing between the Cd and Re sites was observed in the Rietveld refinement. In contrast, splitting of some diffraction peaks was observed at 4 K. For example, the inset of Fig. 1a shows a splitting of the 4 0 0 reflection into two peaks. The low- and high-angle peaks have a two-to-one intensity ratio, indicating a structural change to a tetragonal cell with a $c$-axis length longer than the $a$-axis length. No superlattice reflection was observed at 4 K, which implies that the primitive cell is unchanged through the transition.

The high-temperature face-centred cubic cell was transformed into a low-temperature body-centred tetragonal cell. All reflections in the 4 K pattern are indexed to a tetragonal cell of the space group $I4/m$ with lattice constants of $a = 5.86948(4)$ Å and $c = 8.32632(8)$ Å. This space group is identical to that of the low-temperature structure of the sister compound $Ba_2CaReO_6$ [40]. The structural transition possibly occurs at $T_s = 170$ K, where anomalies in the heat capacity and linear thermal expansion were observed, as shown in Figs. 1b and c.

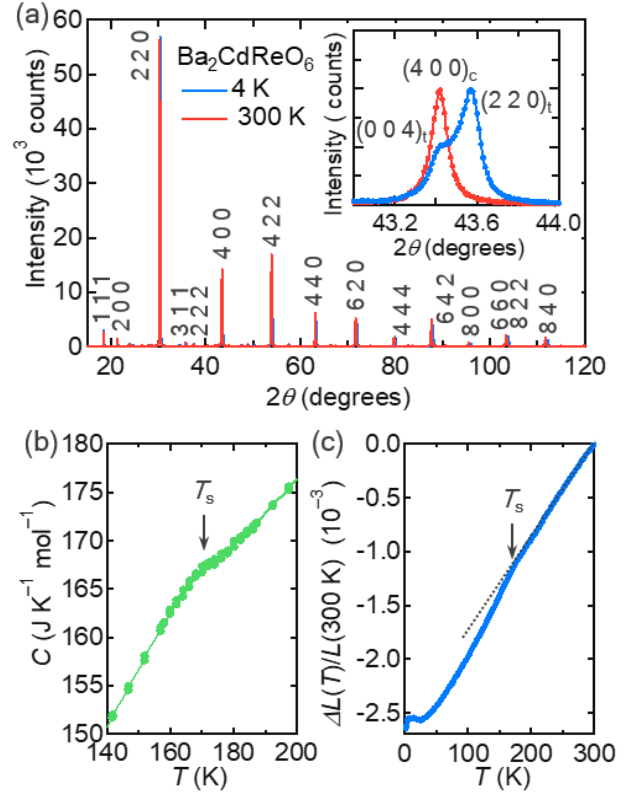

**Figure 1.** (a) Powder XRD patterns at 300 and 4 K for the DP $Ba_2CdReO_6$. The peaks at 300 K are indexed based on a cubic unit cell with $a = 8.3253(1)$ Å in the space group $Fm\text{–}3m$. The inset shows that a single peak of the 4 0 0 reflection of the cubic cell at 300 K splits into two peaks of 0 0 4 and 2 2 0 reflections of the low-temperature tetragonal cell at 4 K. The subscripts t and c denote the index based on the tetragonal and cubic cells, respectively. The temperature dependences of (b) heat capacity and (c) linear thermal expansion $\Delta L(T)/L(300\ \text{K})$ show anomalies in the structural transition at $T_s = 170$ K.

The cubic-to-tetragonal transition from the space group $Fm\text{–}3m$ to $I4/m$ is common to several DPs [42,43]. The origin of this structural transition is considered to be a size mismatch between the constituent ions. In the case of the simple cubic perovskite $ABO_3$, the A–O bond length ideally equals the B–O bond length multiplied by $\sqrt{2}$. Thus, the size mismatch between the A and B cations is described by the Goldschmidt tolerance factor, $t = (r_A + r_O)/\sqrt{2}(r_B + r_O)$, where $r_A$, $r_B$, and $r_O$ are the ionic radii of A, B, and O, respectively [44]. When the A cation size is relatively small, the size mismatch is accommodated by tilting or rotation of the $BO_6$ octahedra. For DPs, the average of the ionic radii of the $B$ and $B'$ cations is used for $r_B$. The tolerance factors of $Ba_2B\text{ReO}_6$ were calculated to be $t = 1.046, 1.041, 0.990$, and $0.979$ for $B$ = Mg, Zn, Cd, and Ca, respectively (Table 1), using the ionic radii from the literature [45]. Since $t$ is smaller than unity for $B$ = Cd and Ca, a structural instability causing tilt or rotation of the octahedra is expected. In fact, Ba$_2$CaReO$_6$ shows a structural transition from a cubic $Fm$–$3m$ structure to a tetragonal $I4/m$ structure at 120 K [40]. In contrast, the Mg and Zn compounds exhibit no such transition. Therefore, it is natural to assume that the same type of structural transition occurs in Ba$_2$CdReO$_6$ at $T_s$ = 170 K. The reason why the Cd compound with $t$ closer to unity has a slightly higher transition temperature than that of the Ca compound is not clear; the structural stability may be in a delicate balance and is influenced by factors other than the ion size mismatch.

**Table 1.** Comparison between the parameters of DP type rhenates with the 5$d^1$ electronic configuration. $a$ is the lattice constant of the cubic $Fm$–$3m$ structure at room temperature. $T_s$ is the temperature of the structural transition from $Fm$–$3m$ to $I4/m$. $T_q$ and $T_m$ refer to the temperatures of quadrupole and magnetic orders, respectively. $M_{sat}$ is the saturation magnetic moment at 2 K, and $\mu_{eff}$ is the effective magnetic moment in the unit of the Bohr magneton obtained from Curie–Weiss fitting for the data measured at a magnetic field of 7 T. The three values for the Mg compound are the single crystal data obtained in the magnetic field along the [100], [110], and [111] directions.

| $B$ in Ba$_2B$ReO$_6$ | Mg | Zn | Cd | Ca |
|---|---|---|---|---|
| Ionic radius of B$^{2+}$ ion (Å) | 0.720 | 0.740 | 0.95 | 1.00 |
| Tolerance factor | 1.046 | 1.041 | 0.990 | 0.979 |
| $a$ (Å) | 8.0802(2) | 8.1148(1) | 8.3253(1) | 8.371(4) |
| $T_s$ (K) | NA | NA | 170 | 120 |
| $T_q$ (K) | 33 | 33? | 25 | NA |
| $T_m$ (K) | 18 | 11 | 12 | 15.4 |
| $M_{sat}$ at 2 K ($\mu_B$) | 0.254, 0.265, 0.307 | 0.1 | 0.211 | NA |
| $\mu_{eff}$ ($\mu_B$) above $T_q$ | 0.678(1), 0.689(1), 0.673(1) | 0.940 | 0.720(1) | 0.744(2) |
| $\mu_{eff}$ ($\mu_B$) between $T_q$ and $T_m$ | 0.459(3), 0.472(4), 0.478(5) | NA | 0.50(1) | NA |
| $\Theta_W$ (K) above $T_q$ | −14.6(2), −15.2(3), −11.2(2) | −66(2) | −15.3(6) | −38.8(6) |
| $\Theta_W$ (K) between $T_q$ and $T_m$ | 20.2(1), 20.0(1), 19.7(2) | NA | 10.3(1) | NA |
| References | [30,36] | [36] | This study | [40] |

### 3.2 Electronic properties

The electric resistivity of Ba$_2$CdReO$_6$ was as high as 0.2 MΩ cm at room temperature. Its temperature dependence follows a thermal activation type, $\rho(T)$ = A exp($\Delta/T$), with an activation energy of $\Delta$ = 0.24 eV, which is comparable to those of other 5$d^1$ DPs: $\Delta$ = 0.17 eV for Ba$_2$MgReO$_6$ [30] and $\Delta$ = 0.26 eV for Ba$_2$CaReO$_6$ [39]. Thus, Ba$_2$CdReO$_6$ is a Mott insulator with a relatively small energy gap.

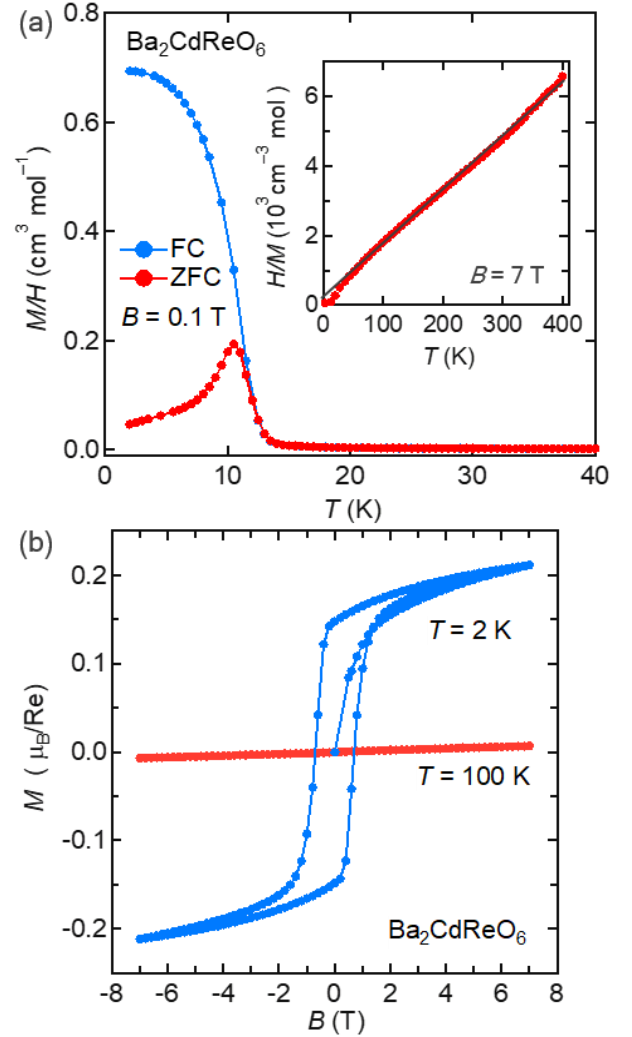

**Figure 2.** (a) Temperature dependence of the magnetic susceptibilities of polycrystalline Ba$_2$CdReO$_6$ measured during heating after zero-field cooling (ZFC), and on cooling in an applied magnetic field of 0.1 T (FC). The inset shows an inverse magnetic susceptibility measured at $B$ = 7 T. The black line represents the Curie–Weiss fit between 100 and 400 K, which yields a Weiss temperature of $\Theta_W$ = −15.3(6) K and an effective magnetic moment of $\mu_{eff}$ = 0.720(1) $\mu_B$. (b) Isothermal magnetizations measured at 100 and 2 K.

Magnetic susceptibility measurements revealed the spin–orbit-entangled electronic state of Ba$_2$CdReO$_6$. The inverse susceptibility, shown in the inset of Fig. 2a, exhibits a linear temperature dependence over a wide temperature range between 50 and 400 K, demonstrating a Curie–Weiss-type magnetism for localised 5$d$ electrons in the paramagnetic state. There is no visible anomaly at the structural transition at $T_s$ =

170 K, which indicates that magnetic interactions between 5$d$ electrons are not affected by the structural transition. A fit to the Curie–Weiss formula, $1/\chi = (T - \Theta_W)/C$, in a temperature range of 100–400 K yields a Weiss temperature of $\Theta_W = -15.3(6)$ K and a Curie constant of $C = 0.0648(1)$ cm$^3$ K mol$^{-1}$. The negative Weiss temperature indicates dominant AF interactions. The effective magnetic moment is calculated to be 0.720(1) $\mu_B$, which is considerably smaller than the 1.73 $\mu_B$ expected for spin-1/2. This large reduction is ascribed to the cancelation of spin and orbital angular momenta coupled antiparallelly by SOIs. Although the magnetic moment should be zero ($M = 2S - L = 0$, for an ideal $J_{eff} = 3/2$ state) non-zero moments have been observed in real materials, which may be owing to hybridisation between the 5$d$ state of the transition metal and the 2$p$ state of oxygen [46,47] as well as the contribution of an excited $J_{eff} = 1/2$ state [13]. The reduced moment of Ba$_2$CdReO$_6$ is comparable to the ~ 0.68 $\mu_B$ of Ba$_2$MgReO$_6$ [30] and ~ 0.60 $\mu_B$ of Ba$_2$NaOsO$_6$ [29], demonstrating that similar spin–orbit-entangled electronic states are realised in these compounds. Note that, even for 5$d$ electron systems, the effective moment is approximately 1.73 $\mu_B$ when the orbital angular momentum is quenched by a low-symmetry crystal field. For example, a quantum magnet Ca$_3$ReO$_5$Cl$_2$ containing Re$^{6+}$ ions in the distorted O$_5$Cl octahedron has an effective moment of 1.59 $\mu_B$ [48,49].

As shown in Fig. 2, the magnetic susceptibility increases steeply below 12 K, which signals a transition to a magnetic-dipole ordered phase. The large increase below 12 K and the difference between the field-cooling and zero-field-cooling curves is owing to the appearance of an FM moment. This is evidenced by the hysteresis in the isothermal magnetisation curve at 2 K (Fig. 2b). The saturation magnetic moment of 0.21 $\mu_B$ at 2 K and 7 T is too small for a ferromagnet, and is comparable to the saturation moments of ~ 0.3$\mu_B$ and 0.25$\mu_B$ reported for the noncollinear AF orders in the sister compounds Ba$_2$MgReO$_6$ [30] and Ba$_2$NaOsO$_6$ [29], respectively. Considering the dominant AF interactions, the magnetic order of Ba$_2$CdReO$_6$ is likely a similar noncollinear AF structure with a large canting angle. This unusual canted AF order has been suggested to be stabilised by the preceding quadrupole order at high temperatures [10,35]. Thus, we carefully examined the physical properties above $T_m$.

### 3.3 Possible quadrupole transition

Figure 3 shows the inverse magnetic susceptibility, heat capacity, and linear thermal expansion data below 50 K. In the heat capacity data, a peak indicative of a second-order phase transition is observed at $T_m = 12$ K, confirming the magnetic phase transition. In addition, there is another peak at $T_q = 25$ K, demonstrating an additional bulk phase transition. The absence of a distinct anomaly in the magnetic susceptibility at $T_q$ suggests that this phase transition is not magnetic in origin. On the other hand, a kink is observed in the linear thermal expansion data, as shown in Fig. 3c. Generally, thermal expansion measurements can sensitively probe a structural change. Thus, the $T_q$ transition must be either a purely structural transition or an electronic transition that is accompanied by a structural change.

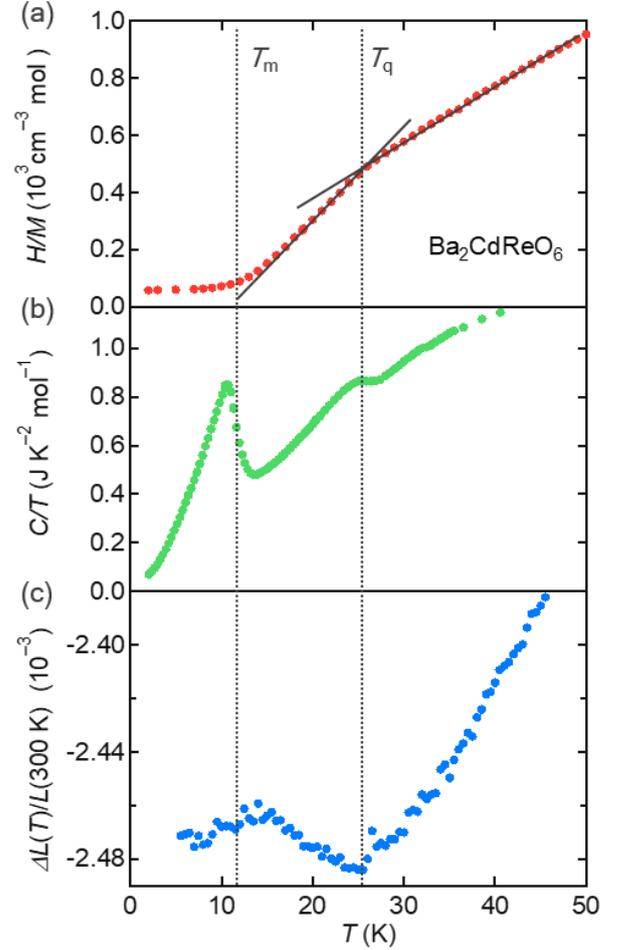

**Figure 3.** Anomalies at $T_q = 25$ K and $T_m = 12$ K observed in the temperature dependences of (a) inverse magnetic susceptibility measured at a magnetic field of 7 T, (b) heat capacity, and (c) linear thermal expansion. The black solid lines in (a) represent the two Curie–Weiss regions across $T_q$.

Although there is no clear anomaly in the magnetic susceptibility, the slope of the inverse susceptibility changes after crossing $T_q$. A Curie–Weiss fit to the narrow temperature range between $T_q$ and $T_m$ gives a positive Weiss temperature of $\Theta_W = 10.3(1)$ K and an effective magnetic moment of $\mu_{eff} = 0.50(1)$ $\mu_B$, while $\Theta_W = -15.3(6)$ K and $\mu_{eff} = 0.720(1)$ $\mu_B$ are obtained for a fit above $T_q$. Thus, the dominant magnetic interactions switch from AF to FM across $T_q$, and the effective magnetic moment is reduced by 70%. These changes must originate from a crucial change in the electronic state at $T_q$. Similar changes were observed in the Mg compound at the quadrupole order transition: upon cooling, the Weiss temperature changed from −15 to 20 K, and the effective moment decreased from 0.68 $\mu_B$ to 0.47 $\mu_B$ [30]. Therefore, the $T_q$ transition of the Cd compound is likely to be due to a quadrupole order. The reduction of the effective moment has been attributed to the lifting of the degeneracy of the $J_{eff} = 3/2$ quartet by the quadrupole order. In fact, theory predicts a reduction of magnetic moment by a factor of $(3/5)^{1/2}$ (~ 77%) by a quadrupole ordering [13], which is in good agreement with the experimental observations. Once a quadrupole order sets in, the interactions between the 5$d$ electrons are modified and become anisotropic, which may result in the sign change of the Weiss temperatures, as observed in these DPs.

The $J_{eff} = 3/2$ quartet state has fifteen possible active moments, including three from magnetic dipoles, five from electric quadrupoles, and seven from magnetic octupoles [10,50]. Among them, the dipole and octupole moments can give rise to magnetic orders, while the quadrupole moments can form an order of electronic charges. Only a quadrupole moment, which is an anisotropic distribution of charges around the nuclei, can induce lattice distortion through electron–phonon interactions. However, the induced lattice distortion must be small because coupling to the lattice is generally weak. In fact, the structural change induced by the quadrupole order in Ba$_2$MgReO$_6$ is extremely small [32]. For Ba$_2$CdReO$_6$, our laboratory XRD measurements failed to detect a structural change at $T_q$ = 25 K, but the linear thermal expansion in Fig. 3c shows a clear kink: from 50 K, it decreases upon cooling and starts to increase below $T_q$. This negative thermal expansion is probably attributed to the development of an electronic order. Therefore, it is plausible that the quadrupole moments order at $T_q$. The linear thermal expansion decreases again below $T_m$, implying that the negative thermal expansion is suppressed by the magnetic dipole order. It is likely that the quadrupole moments themselves are reduced upon cooling below $T_m$ with the relative dipole components increasing.

## 4. Discussion

### 4.1 Chemical trend

The physical properties of Ba$_2$CdReO$_6$ are compared with those of the related DPs Ba$_2B$ReO$_6$ in Table 1. With an increase in the size of cations at the $B$ site, the lattice expands systematically. The crystal structures of $B$ = Mg and Zn with tolerance factors larger than one remain cubic at low temperatures and pseudo-cubic with very small tetragonal deformation below $T_q$ for $B$ = Mg. In contrast, cubic-to-tetragonal structural transitions accompanied by a rotation of the $B$O$_6$ octahedra occur for DPs with tolerance factors smaller than unity at $T_s$ = 170 and 120 K for $B$ = Cd and Ca, respectively. On the other hand, from an electronic point of view, the Mg, Zn, and Cd compounds share a common feature at low temperatures, while the Ca compound is different: the former exhibit quadrupole ordering at $T_q$ followed by canted AF ordering with large FM moments below $T_m$, whereas the latter shows a single transition to the collinear AF order [40]. It is also noted that the properties of $B$ = Mg and Cd are significantly similar to each other: the saturation moment of the magnetically ordered phase is approximately 0.2 $\mu_B$, and two Curie–Weiss regions exist in the inverse magnetic susceptibilities across $T_q$. It is remarkable that the Cd compound shows a cubic-to-tetragonal transition at high temperatures, similar to the Ca compound, and successive transitions to the quadrupole and magnetic orders, like the Mg compound.

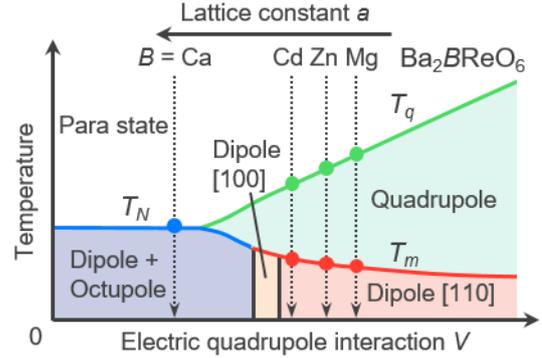

**Figure 4.** Schematic mean-field phase diagram for the $d^1$ DP with the $J_{eff} = 3/2$ quartet state depicted as a function of the electric quadrupole interaction $V$ for $J'/J = 0.2$. 'Dipole [1 1 0]' and 'Dipole [1 0 0]' refer to the two-sublattice noncollinear dipole orders with the net FM moments pointing along the [1 1 0] and [1 0 0] axes, respectively. 'Quadrupole' and 'Dipole + Octupole' refer to multipole orders of electric quadrupoles and orders of magnetic dipoles and octupoles, respectively. The DPs Ba$_2B$ReO$_6$ ($B$ = Mg, Zn, Cd, and Ca) are approximately located at the positions marked by the arrows. The lattice constants increase to the left with decreasing $V$. The 'Dipole [1 1 0]' and 'Dipole [1 0 0]' orders are not distinct as they differ only in the direction of the net moment. The Cd compound may be located in the 'Dipole [1 0 0]' region.

It is expected that a cubic-to-tetragonal structural transition at high temperatures lifts the multipolar degeneracy and suppresses a multipolar order at low temperatures since a structural transition often induces a distortion of the $B$O$_6$ octahedron that quenches the orbital angular moment of $d$ electrons. However, this was not the case for the Cd compound. The $J_{eff} = 3/2$ electronic state is preserved even in the tetragonal phase, as evidenced by the reduced effective magnetic moment below $T_s$. Similarly, for the Ca compound, the inverse magnetic susceptibility does not show a visible change at the structural transition[40], reflecting a negligible change in the spin-orbit entangled electronic state. The cubic-to-tetragonal transitions are mainly caused by the rotations of the undeformed ReO$_6$ octahedra. Therefore, the degeneracy of the $J_{eff} = 3/2$ quartet is not lifted in the tetragonal phase. This may also be the case for other tetragonally distorted 5$d^1$ DPs:

considerably reduced effective magnetic moments have been reported for $Sr_2CaReO_6$ (0.74 $\mu_B$) [40] and $Sr_2MgReO_6$ (0.80 $\mu_B$) [51]. Thus, multipolar orders derived from the spin–orbit-entangled $J_{eff}$ = 3/2 state should occur not only in cubic DPs but also in tetragonal DPs.

*4.2 Electronic phase diagram*

Finally, the variation of the ground states in this DP family is discussed based on a phase diagram obtained by the mean-field theory. G. Chen et al. constructed a phase diagram for spin–orbit-entangled DPs with the $d^1$ electronic configuration. They introduced three types of interactions [10,12]: the nearest-neighbour AF exchange $J$, the nearest-neighbour FM exchange $J'$, and the electric quadrupole interaction $V$. Figure 4 shows a schematic phase diagram as a function of $V$ at a fixed $J'/J$ of 0.2. In a region with relatively large $J'$ and $V$, there is a noncollinear magnetic order with [1 1 0] anisotropy (Dipole [1 1 0]) as the ground state and a quadrupole ordered phase at high temperature, which is the precise observation in case of the Mg compound. It is emphasised that the quadrupole order should exist above Dipole [1 1 0] because the former is required to stabilise the latter. On the other hand, an AF phase accompanied by an octupole order (Dipole + Octupole) is located in a region with a small $V$, which may be the case for the Ca compound. There is no other symmetry breaking above the octupole order. It is also noted that this complex order looks like a simple dipole order because the octupole component is invisible in conventional magnetic measurements. In addition, there is another noncollinear magnetic order with [1 0 0] anisotropy (Dipole [1 0 0]) between the two ground states.

Considering the correspondence between the theoretical phase diagram and the experimental results on the DPs, it is plausible that the Mg, Zn, and Cd compounds are located in the quadrupole order regime on the right of the phase diagram and the Ca compound in the octupole order regime on the left. The observed variation in the ground states is possibly caused by the subtle variations of $J'$ and $V$, depending on the lattice constants. The FM interactions $J'$ seem to be small based on the AF Weiss temperatures for the DPs. The electronic quadrupole interactions $V$ should depend on the lattice constant: the larger the lattice constant, the smaller the value of $V$, which is given by $9\sqrt{2}Q^2/a^5$, where $Q$ is the magnitude of the electric quadrupole moment. $Q$ depends on the degree of hybridisation between the rhenium $5d$ states and oxygen $2p$ states, which is smaller for a longer TM–O distance. Thus, $Q$ is smaller for larger $a$. For a fixed $Q$, the difference in the lattice constants results in a 16% smaller $V$ for $B$ = Ca than that for $B$ = Mg. Consequently, as illustrated in Fig. 4, the Ca, Cd, Zn, and Mg compounds are located in the right of the phase diagram with decreasing $a$ and increasing $V$. Therefore, the variations in the properties of the DPs are clearly explained by this phase diagram.

*4.3 Future direction*

$Ba_2BReO_6$ provides an ideal material platform to experimentally investigate the physics of multipolar degrees of freedom in spin–orbit-entangled $d$ electrons. In a future study, it would be interesting to observe a switching between the ground states in the solid solutions of $Ba_2CdReO_6$ and $Ba_2CaReO_6$. Furthermore, the ground state of $Ba_2CaReO_6$ may be tuned by applying pressure, which will reduce the lattice constant and enhance $V$, so that the phase diagram can be studied continuously. More sophisticated experiments, such as inelastic neutron scattering experiments, would reveal the excitation spectra, which would provide us with important information on the multipolar ground states of $5d$ electron systems.

## 5. Summary

The physical properties of $5d^1$ DP $Ba_2CdReO_6$ were investigated. This compound undergoes a structural transition from cubic to tetragonal at $T_s$ = 170 K, which possibly originates from the size mismatch of the metal ions. A spin–orbit-entangled $J_{eff}$ = 3/2 state is confirmed by the reduced effective magnetic moment of 0.72 $\mu_B$, which is constant below $T_s$. At low temperatures, successive transitions to a possible quadrupole order and a canted AF order with a large FM moment appear at $T_q$ = 25 K and $T_m$ = 12 K, respectively. Based on a previous work on related DP compounds, the chemical trend for $Ba_2BReO_6$ ($B$ = Mg, Zn, Cd, and Ca) was discussed. A similar cubic-to-tetragonal transition is observed for $B$ = Ca, while the low-temperature successive transitions were similar to those for $B$ = Mg and Zn; the Ca compound showed a different collinear AF ground state. These variations were explained by a theoretical phase diagram as a function of the electric quadrupole interaction, $V$. It is suggested that the chemical pressure from the $B$ cations varies $V$ so that the low-temperature states change between an 'Dipole + Octupole' order for $B$ = Ca and a quadrupole order followed by a dipole order for $B$ = Mg, Zn, and Cd. The $5d^1$ DP $Ba_2BReO_6$ compounds provide us with a suitable platform for systematically studying the rich physics of multipoles of $5d$ electrons.

**Acknowledgements**

The authors are grateful to T. Yajima for the technical support during low-temperature XRD measurements. This work was partly supported by Japan Society for the Promotion of Science (JSPS) KAKENHI Grant Number JP18H04308 (J-Physics), JP19H04688, JP20H05150 (Quantum Liquid Crystals) and JP20H01858 and by the Core-to-Core Program (A) Advanced Research Networks.